\documentclass{article}
\usepackage{graphicx}
\usepackage{amsmath}
\usepackage{amsfonts}
\usepackage{amssymb}
\usepackage{float}
\usepackage{tabularx}
\usepackage[left=2cm, right=2cm]{geometry}
\usepackage{subcaption}
\usepackage{authblk}

\title{Closure Term Estimation in Spatiotemporal Models of Dynamical Systems}
\author[1,2]{Eric Crislip}
\author[3]{Mohammad Khalil}
\author[2]{Teresa Portone}
\author[1]{Oksana Chkrebtii}
\author[2]{Kyle Neal}
\affil[1]{Department of Statistics, The Ohio State University, Columbus, Ohio, 43210, USA}
\affil[2]{Sandia National Laboratories, Albuquerque, New Mexico, 87185, USA}
\affil[3]{Sandia National Laboratories, Livermore, California, 94551, USA}
\date{}

\begin{document}

\maketitle

\section*{Abstract}
Closure  modeling — the statistical modeling of missing dynamics in the natural sciences and engineering — is a growing and active area of research. Existing methods for closure modeling are often computationally prohibitive, lack uncertainty quantification, or require noise-free observations of the temporal derivatives over the system state. We propose a novel, computationally efficient approach for the modeling and estimation of closure terms over the spatiotemporal domain that provides uncertainty quantification and is effective even when the observations of the system state are sparse or contain moderate levels of noise. The efficacy of our approach is demonstrated in both one and two spatial dimensions through numerical experiments using the Fisher-KPP reaction-diffusion equation and the advection-diffusion equation as exemplars. 

\section{Introduction}
Physics-based models for engineering and the sciences, often taking the form of ordinary or partial differential equations (ODE, PDE), may inadequately represent real-world phenomena due to missing, unknown, or simplified system dynamics. The field of scientific machine learning (SciML) has recently emerged to ameliorate predictive deficiencies or inferential biases in theoretically-derived physics-based models by leveraging tools from machine learning and real-world observations. Popular SciML approaches for improving predictions from physics-based models have included directly modeling the discrepancy between physics-based model output and observed measurements \cite{Kennedy}, incorporating governing equations as soft constraints in data-driven machine learning models \cite{RAISSI2019686,Yang_2021}, and more recently, directly learning missing terms from the governing equations themselves \cite{subramanian2019error, rackauckas2021universaldifferentialequationsscientific}. The last approach, henceforth referred to as \emph{closure modeling} \cite{sanderse2024scientificmachinelearningclosure}, has several distinct advantages, including greater interpretability, the potential to improve predictions for unmeasured quantities of interest, and the ability to impose physical constraints in the machine learning-corrected physics-based model. 

The use of data assimilation tools has attracted some recent attention for estimating the closure term from observed system responses \cite{subramanian2019error,garg2022physics,KASHYAP2024111474,neal2024investigating}. Data assimilation is the process of synthesizing scientific or engineering physics-based models with real-world data to create estimates of the unknown system states that are evolving over time. The central idea of sequential data assimilation for state estimation is to cyclically \emph{forecast} future values of the state based on an understanding of system dynamics and previously obtained observations, then update these forecasts by an \emph{analysis} step when new observations become available. If desired, past predictions may be updated with current data to obtain smoother estimates. The Bayesian interpretation of data assimilation represents the forecast and analysis cycle probabilistically through the forecast and analysis probability distributions, which inherently allows for uncertainty quantification \cite{sullivan2015introduction}. When new data arrive, the forecast distribution may be updated to the analysis distribution by applying Bayes' law, while the distributions for past states can be updated to smoothing distributions via backwards recursion \cite{godsill2004monte}.

Among the most widely studied methods for data assimilation is the Kalman filter, which provides exact closed-form expressions for the analysis and forecast distributions when both the dynamics and observation process are linear and Gaussian \cite{kalman1960new,evensen2006data}. The assumption of linear dynamics makes the naive Kalman filter unsuitable for nonlinear dynamical systems that are often encountered in practice. A popular extension of the Kalman filter that allows for nonlinear evolution models is the ensemble Kalman filter (EnKF), which represents forecast and analysis probability distributions using a set of Monte Carlo particles known as ensemble members \cite{evensen2009ensemble}. In the EnKF, the ensemble members evolve forward in time to provide an approximation of the forecast distribution, which are then updated to an approximation of the analysis distribution using the closed-form expression of the Kalman filter. These closed-form expressions are derived from the Kalman filter's assumption of joint Gaussianity for the state and observations, which generally does not hold for nonlinear evolution models, yet the approximations induced in the forecast and analysis distributions by the closed-form analysis step of the EnKF often provide reasonable results owing the correspondence with Bayesian linear estimation theory \cite{hartigan1969linear, goldstein2007bayes,west2006bayesian}. Furthermore, the EnKF is fairly robust to deviations from the Gaussianity assumption for the data and extensions to the EnKF exist to accommodate strongly non-Gaussian observations. The EnKF is often combined with \emph{state augmentation}, where the state vector is augmented with model parameters from the governing equations \cite{Anderson:2001,evensen2009ensemble}. State augmentation allows the model parameters to be inferred from the data through the filtering process as part of the analysis step. Parameter learning via state augmentation has been widely successful for many applications, though it has been shown to fail for certain variance parameters \cite{Stroud:2010}. Our proposed method utilizes state augmentation to provide computationally efficient estimates of the closure term, with any auxiliary parameters estimated by maximum marginal likelihood (empirical Bayes under some conditions).

While data assimilation tools have previously been applied in closure modeling, the aim of this work is to extend previous work on estimating closure terms with data assimilation for systems described by ODEs to PDEs and other spatiotemporal models \cite{subramanian2019error,garg2022physics,KASHYAP2024111474,neal2024investigating}. To facilitate the use of data assimilation in capturing the unknown spatial closure term, we use basis expansion methods for function approximation \cite{hastie2009elements}. As a means of dimensionality reduction, basis expansion of the spatial closure component increases computational efficiency and reduces the variability in the closure term estimates relative to closure estimation at each spatial point over a discrete spatial grid, as well as mitigate potential adverse effects from spurious correlations in the EnKF procedure. Furthermore, basis expansion defines a desired level of spatial smoothness in the closure term estimates. Basis expansion methods have a long and rich history in the fields of approximation theory and nonlinear regression. We will give particular attention to Gaussian kernel bases, which are known to be universal approximators for continuous functions \cite{buhmann2000radial, du2006radial,park1991universal}, and B-splines, which are are a particularly popular choice of basis for univariate regression \cite{de1978practical,ramsay2002applied}.

Current work for closure modeling in physics-based models governed by PDEs primarily focuses on the setting where observations of the temporal derivatives are assumed to be available \cite{PDEfind} or attempts to estimate a parametric model for the functional relationship between the discretized state and closure term through an expensive calibration procedure \cite{rackauckas2021universaldifferentialequationsscientific}. Our method for closure term estimation can be combined with machine learning tools to model the functional relationship between the state and closure term as part of a two-staged approach \cite{subramanian2023probabilistic} or on its own as a standalone means of model validation and diagnostics, sensitivity analysis, and diagnostics. The two-stage approach is as follows: in the first stage, the value of the system state and closure term are estimated by data assimilation using our methodology. In the second stage, the estimates of the system state and closure term are collected as input-output pairs and used as ``data" to train a machine learning model for the functional form of the closure. We conjecture that the two-staged approach to closure modeling will have several advantages. Namely, the decoupling of the closure term estimation and training of the machine learning closure model ensures that only a fixed number of runs of the expensive physics-based model will be utilized. Furthermore, data assimilation, when viewed from a Bayesian perspective, comes with built-in uncertainty quantification, which may be propagated through the machine learning mode. However, the focus of this paper is on the estimation of the closure term, and subsequent second-stage modeling will be the focus of future work.

This paper is organized as follows. Section 2 reviews the ensemble Kalman filter and describes our proposed methodology. Section 3 details numerical experiments on (1) the 1D Fisher-KPP equation, (2) the 1D advection-diffusion equation, (3) the 2D Fisher-KPP equation, and (4) the 2D advection-diffusion equation. Finally, the conclusions are presented in Section 4.

\section{Methodology}

Next, we describe the proposed methodology for closure term estimation in spatiotemporal models. Section 2.1 introduces data assimilation and some required notation. Section 2.2 provides a review of the stochastic ensemble Kalman filter for state estimation. Section 2.3 outlines existing work on closure term estimation using the EnKF for systems based on ordinary differential equations. In Section 2.4, we propose an extension of the EnKF-based closure term estimation methodology to PDE models. Section 2.5 provides a discussion on the choice of spatial basis functions used in our proposed methodology, while Section 2.6 details a maximum likelihood selection criterion for estimating the hyperparameters required to implement our method in practice.

\subsection{Review of data assimilation for state estimation}
We begin with a review of the data assimilation for state estimation and some notation. Suppose we collect observations of the state at times $0 \leq t_1 < t_2 < \ldots < t_{N_o} \leq T$. Let us denote the spatially discretized state vector at time $t \in [0,T]$ by $\mathbf{u}_t \in \mathbb{R}^{N_x}$. The dimension $N_x$ may refer to a finite number of state variables or a vectorized, discrete representation of one or more fields. If we are interested in the scalar field $u: \Omega \times [0,T] \rightarrow \mathbb{R}$ and let $\mathbf{x}_1, \mathbf{x}_2, \ldots, \mathbf{x}_{N_x}$ be a set of spatial grid points over the spatial domain $\Omega \subseteq \mathbb{R}^p$, then 
\begin{equation}
\mathbf{u}_t = \begin{bmatrix}
u(\mathbf{x}_1,t), u(\mathbf{x}_2,t), \ldots, u(\mathbf{x}_{N_x},t)
\end{bmatrix}^\top. 
\end{equation}

Now suppose the data at time $t_i$, $\mathbf{y}_{t_i} \in \mathbb{R}^{N_{t_i}}$, consists of $N_{t_i}$ measurements and is obtained from the observation model $\mathbf{y}_{t_i} = \mathcal{H}_{t_i} (\mathbf{u}_{t_i},\boldsymbol \epsilon_t)$ with measurement noise $\boldsymbol \epsilon_{t_i}$. In the context of the ensemble Kalman filter (EnKF), the observation model is typically assumed to be linear with additive Gaussian noise,
\begin{align}
\label{eq:obsmodel}
    \mathbf{y}_{t_i} &= \mathbf{H}_{t_i} \mathbf{u}_{t_i} + \boldsymbol \epsilon_{t_i}, \quad
    \boldsymbol \epsilon_{t_i} \sim \mathcal{N}(\mathbf{0},\mathbf{R}_{t_i}).
\end{align}
Here $\mathbf{H}_{t_i} \in \mathbb{R}^{N_{t_i} \times N_x}$ is a linear observation matrix and $\{\boldsymbol \epsilon_{t_i}\}_{i=1}^{N_o}$ are temporally independent random variables. Modifications to the EnKF exist that can accommodate nonlinear observation operators $\mathcal{H}_{t_i}(\cdot,\cdot)$ \cite{evensen2009ensemble}, but we will assume the observation process is linear in this work. For simplicity, we will also assume a diagonal observation error covariance matrix $\mathbf{R}_{t_i} = \gamma \mathbf{I}_{N_{t_i}}$ with known noise parameter $\gamma$ in all of our numerical experiments. This assumption corresponds to spatially independent observation errors, but the method is still applicable when the observations are spatially correlated.

The discrete-time physics-based model for the state $u$ may be represented by the equations
\begin{equation}
\mathbf{u}_{t_{i+1}} = \mathcal{M}_{t_i}(\mathbf{u}_{t_i}), \quad i = 1, \ldots, N_o,
\end{equation}
in which the function $\mathcal{M}_{t_i}: \mathbb{R}^{N_x} \rightarrow \mathbb{R}^{N_x}$ is often referred to as the \emph{evolution model} that describes the time-evolution of the state and could be deterministic or stochastic. Implicitly, we assume future values of the state only depend on the state's current value, i.e. the process is first-order Markovian. For example, $\mathcal{M}_{t_i}$ could represent the flow map for a discretized system of deterministic PDEs.

The Bayesian interpretation of data assimilation combines the evolution model and observation model to obtain probabilistic representations of unknown quantities of interest. Two primary objectives under this paradigm are to obtain the \emph{forecast} distribution with density
\begin{equation}
    p(\mathbf{u}_{t} \mid \mathbf{y}_{t_1}, \mathbf{y}_{t_2}, \ldots, \mathbf{y}_{t_{i}}),
\end{equation}
which represents degrees of belief about the state at a future time $t > t_i$ given past data, and the \emph{analysis} distribution with density
\begin{equation}  p(\mathbf{u}_{t_i} \mid \mathbf{y}_{t_1}, \mathbf{y}_{t_2}, \ldots, \mathbf{y}_{t_i}),
\end{equation}
which represents belief about the system state at the time an observation is collected, given all the data available at the time of analysis. These distributions can be constructed from the evolution and observation models after eliciting a prior distribution $p(\mathbf{u}_0)$ on the state at the initial time, which should reflect \emph{a priori} knowledge about the initial system state and may be chosen to obey known physical constraints. If the initial condition is known, then this prior would be a point mass distribution on the initial value. The forecast and analysis distributions follow from the prior distribution through the repeated application of Bayes' law.

When the evolution model is nonlinear, the forecast and analysis distributions are typically unavailable in closed-form \cite{WIKLE20071}, but may be estimated in some cases from a Monte Carlo sample. These methods generate individual particles, or \emph{ensemble members}, that form an approximate sample from the required distributions. For many complex physical systems, characterization of the full forecast and analysis probability distributions is computationally infeasible, as typical sequential Monte Carlo methods such as particle filters may require sample sizes many orders of magnitude greater than the dimension of the associated state space \cite{Bengtsson_2008}. An alternative to Monte Carlo methods that approximate the full forecast and analysis distributions is the ensemble Kalman filter, which merely represents the aforementioned probability distributions by their mean and covariance. The ensemble mean of the ensemble Kalman filter has enjoyed great success as an estimator of the system state in many applications with relatively small ensemble sizes \cite{AssessingthePerformanceoftheEnsembleKalmanFilterforLandSurfaceDataAssimilation,gillijns2006ensemble}, thus it is our preferred technique for data assimilation.

\subsection{The stochastic ensemble Kalman filter for state estimation}

With the evolution and observation models and the prior distribution on the state at the initial time specified, we may present an algorithmic treatment of the ensemble Kalman filter. The EnKF begins by initializing ensemble members as Monte Carlo realizations from the prior on the state at the initial time
\begin{equation}
    \mathbf{u}_{0,\ell} \overset{iid}{\sim} p(\mathbf{u}_0), \quad \ell = 1, 2, \ldots, N_{ens}.
\end{equation}
Each ensemble member is then pushed through the evolution model $\mathbf{u}^f_{t_1,\ell} = \mathcal{M}_0(\mathbf{u}^f_{0,\ell})$ to obtain samples from the forecast distribution at time $t_1$. To update the forecast ensemble at time $t_i$ to the analysis ensemble at time $t_i$ using the available data $\mathbf{y}_{t_i}$, the forecast ensemble mean and covariance must be first estimated by
\begin{align}
    \hat{\mathbf{m}}_{t_i}^f &= \frac{1}{N_{ens}} \sum_{\ell=1}^{N_{ens}} \mathbf{u}^f_{{t_i},\ell}, \label{eq:analysis_mean}\\
    \widehat{\mathbf{P}}_{t_i}^f &= \frac{1}{N_{ens}-1} \sum_{\ell=1}^{N_{ens}} (\mathbf{u}^f_{t_i,\ell} - \hat{\mathbf{m}}_{t_i}^f)^T (\mathbf{u}^f_{t_i,\ell} - \hat{\mathbf{m}}_{t_i}^f). \label{eq:analysis_cov}
\end{align}

Each ensemble member is then updated from a sample of the forecast distribution at time $t_i$ to a sample from the (approximate) analysis distribution at time $t_i$ according to
\begin{equation}
    \mathbf{u}^a_{t_i,\ell} = \hat{\mathbf{K}}_{t_i} (\mathbf{y}_{t_i} - \mathbf{H}_{t_i} \mathbf{u}^f_{t_i,\ell} + \mathbf{e}_{t_i,\ell}), \quad \ell = 1, \ldots, N_{ens},
\end{equation}
where $\mathbf{e}_{t_i,\ell} \overset{iid}{\sim} \mathcal{N}(\mathbf{0},\mathbf{R}_{t_i})$ are \emph{measurement perturbations} and $\hat{\mathbf{K}}_{t_i}$ is the \emph{Kalman gain} given by
\begin{equation}
\label{eq:kalman_gain}
\hat{\mathbf{K}}_{t_i} = \widehat{\mathbf{P}}_{t_i}^f \mathbf{H}^T_{t_i}(\mathbf{H}_{t_i} \widehat{\mathbf{P}}_{t_i}^f \mathbf{H}^T_{t_i} + \mathbf{R}_{t_i})^{-1}.
\end{equation}
When the forecast distribution is Gaussian (e.g. the assumptions of the Kalman filter hold) and the forecast mean $\mathbf{m}^f_{t_i}$ and covariance $\mathbf{P}^f_{t_i}$ are known, the analysis step updates the forecast distribution ensemble exactly to samples from the analysis distribution 
\begin{equation}
\label{eq:kalman_analysis_distr}
    \mathcal{N}\Big(\mathbf{m}_{t_i} + \mathbf{K}_{t_i}(\mathbf{y}_{t_i}-\mathbf{H}_{t_i}\mathbf{m}_{t_i}),(\mathbf{I}_{N_x}-\mathbf{K}_{t_i}\mathbf{H}_{t_i})\mathbf{P}_{t_i}\Big),
\end{equation}
where the optimal Kalman gain $\mathbf{K}_{t_i}$ is found by using the true parameters in equation (\ref{eq:kalman_gain}). Otherwise, the analysis step in the EnKF can be interpreted as falsely assuming the forecast distribution is Gaussian and approximating the true analysis distribution with (\ref{eq:kalman_analysis_distr}), which may also induce approximations in subsequent forecast distributions. Alternatively, one can view the analysis step of the EnKF through the lens of Bayesian linear estimation theory \cite{hartigan1969linear}, which does not require distributional assumptions for the forecast distribution beyond the mean and covariance. However, both interpretations of the EnKF are only ever approximate due to the Monte Carlo error in equations (\ref{eq:analysis_mean}-\ref{eq:analysis_cov}).

The addition of measurement perturbations as part of the analysis step is the distinguishing feature of the stochastic ensemble Kalman filter, as it ensures the ensemble variance is not underestimated \cite{burgers1998analysis, houtekamer1998data}. Deterministic variants of the ensemble Kalman filter exist and circumvent the need for measurement perturbations, but these methods may be more sensitive to deviations from Gaussianity in the forecast distributions than their stochastic counterparts \cite{lei2010comparison}.

After the ensemble is updated to the analysis distribution, we can then move forward in time through the evolution model and obtain the forecast ensemble 
\begin{equation}
\mathbf{u}_{t_{i+1},\ell}^f = \mathcal{M}_{t_i}(\mathbf{u}_{t_i,\ell}^a), \quad \quad \ell = 1, 2, \ldots, N_{ens}.
\end{equation}
If desired, the evolution model may be applied to the ensemble members repeatedly to obtain forecasts in-between observation times. The algorithm proceeds iteratively, alternating between forecast and analysis steps as new data arrives.

\subsection{EnKF closure estimation for systems based on ODEs}

We will now show how the ensemble Kalman filter can be modified to estimate the missing dynamics of the governing equation for simple models with a single state variable, as in \cite{subramanian2019error,neal2024investigating}. Suppose our governing equation is
\begin{equation}
    \frac{d}{dt} u(t) = F\big(u(t)\big) + \phi\big(u(t)\big),
\end{equation}
where $F: \mathbb{R} \to \mathbb{R}$ represents our incomplete understanding of the system and $\phi: \mathbb{R} \to \mathbb{R}$ is the unknown system dynamics. We assume this system admits a unique solution. The additive nature of the unknown dynamics is not a restrictive assumption, as for any $F$, we we have that $\phi(u(t)) = \frac{d}{dt}u(t) - F(u(t))$. Additive model corrections are commonly found in the literature, see e.g. \cite{subramanian2019error,rackauckas2021universaldifferentialequationsscientific,sanderse2024scientificmachinelearningclosure}.

Let $\xi_t := (\phi \circ u)(t) $ be the true value of the closure term at a given time $t$. Since we do not have access to the missing dynamics $\phi(\cdot)$, we represent our uncertainty about $\xi_t$ with the prior probability distribution
\begin{equation}
    \xi_t \overset{\mathcal{D}}{=} \sigma W(t),
\end{equation}
where $W(t)$ is a Brownian motion, $\sigma$ is a scale parameter, and $\overset{\mathcal{D}}{=}$ denotes equality in distribution. This probability model describes our belief about the time-evolution of the closure term and may be used in the context of data assimilation to form a forecast distribution for $\xi_t$. Brownian motion can be thought of as a continuous-time analogue to a symmetric random walk, hence this model may be interpreted as expressing ignorance about the true evolution of the closure term. Moreover, the increasing variance of Brownian motion ensures that even if the value of the closure term is known at a given point in time, uncertainty about the closure term will increase with the distance from that time point. However, standard Brownian motion starts at a deterministic initial value of $W(0) = 0$, which corresponds to complete certainty in the absence of missing dynamics at time $t = 0$, and therefore may be an unreasonable modeling choice. Instead, we may choose to work with a translated Brownian motion $\tilde{W}(t) = W(t) + \mu$, which will allow us to encode our belief about the initial missing dynamics into a prior distribution with density $p(\mu)$.

With a model for the time-evolution of the closure term established, we may proceed to estimate $\xi_t$ using past and current measurements of the state through the ensemble Kalman filter and the method of state augmentation. That is, we form a new state vector $\mathbf{z}_{t_i} = \begin{bmatrix}
    u_{t_i} \\ 
    \xi_{t_i}
\end{bmatrix}$ to be used as the state in the EnKF. If our original observation model is
\begin{equation}
    y_{t_i} = h_{t_i} u_{t_i} + \epsilon_{t_i}, \quad \epsilon_{t_i} \sim \mathcal{N}(0,\gamma^2),
\end{equation}
then our augmented state vector can be associated with an augmented linear observation matrix $\mathbf{G}_{t_i} = \begin{bmatrix}
    h_{t_i} & 0
    \end{bmatrix}$ to obtain the equivalent observation model
\begin{equation}
    y_{t_i} = \mathbf{G}_{t_i} \mathbf{z}_{t_i} + \epsilon_{t_i}, \quad \epsilon_{t_i} \sim \mathcal{N}(0,\gamma^2).
\end{equation}
The discrete-time evolution model $\mathcal{M}^z_{t_i}(\mathbf{z}_{t_i})$ can be found by solving the system of stochastic differential equations,
\begin{equation}
\begin{aligned}
\label{eq:ode-sde}
du &= \big(F(u) + \xi_t\big) dt, \\
d \xi_t &= \sigma dW.
\end{aligned}
\end{equation}
Samples from the approximate solution to (\ref{eq:ode-sde}) can be obtained by using a stochastic solver, e.g. the Euler-Maruyama method.

Given the (augmented) observation model and evolution model, the EnKF proceeds exactly as outlined in Section 2.1 and returns approximate samples for the forecast and analysis distributions over $\xi_t$.

\subsection{EnKF closure estimation for systems based on PDEs}
The contribution of this work is to extend previous closure estimation methodology from scalar problems to the setting where the closure term is an entire spatial field. Suppose our governing equation is
\begin{equation}
    \partial_t u(\mathbf{x},t) = \mathcal{F}(u)(\mathbf{x},t) + \phi(u)(\mathbf{x},t)
\end{equation}
where $\mathcal{F}(u)$ represents our incomplete understanding of the system and $\phi(u)$ is the missing system dynamics. At a given time $t \in [0,T]$, the closure term $\xi_t(\cdot) := (\phi \circ u)(\cdot,t)$ is no longer a single real number, but is now an entire scalar field over the spatial domain $\Omega \subseteq \mathbb{R}^p$. In many cases, a naive application of the previous approach to closure term estimation is inadequate, as an independent treatment of the closure at each spatial point would produce everywhere discontinuous closure terms, that may suffer from high uncertainty in the analysis distribution. Furthermore, a discrete representation of the closure term that is allowed to grow with the dimension of the state space may quickly become computationally prohibitive in high-dimensional models.

To circumvent the potential pitfalls of the naive approach, we propose reducing the infinite-dimensional estimation problem of the spatial closure term to the estimation of a finite set of basis expansion coefficients. Analogously to the scalar estimation setting, we can represent our uncertainty about $\xi_t$ with the prior probability distribution
\begin{equation}
\label{eq:forecast_closure}
     \xi_t(\mathbf{x}) \overset{\mathcal{D}}{=} \sum_{k=0}^M \sigma_k \eta_{k}(t) b_k(\mathbf{x}),
\end{equation}
where $\{\eta_{k}\}_{k=0}^M$ are independent Brownian motion processes, $\{\sigma_k\}_{k=0}^M$ are scale parameters, $\{b_k\}_{k=0}^M$ are a set of basis functions over the spatial domain, and $\overset{\mathcal{D}}{=}$ denotes equality in distribution. Our probability model for $\xi$ is separated into two distinct components: a temporal component determined by the coefficients $\eta_k$ and a spatial component determined by the choice of bases $b_k$. This space-time decoupling allows for a parsimonious representation of the closure term, ensuring that only a select few coefficients be estimated at each analysis step. Similarly to the scalar setting, when viewed as a continuous analogue to the random walk, the Brownian motion distributions of the coefficients express uncertainty about the temporal evolution of the closure term. The choice of basis, however, encodes explicit knowledge regarding the spatial structure of the underlying closure field variable and should have sufficient flexibility to adequately model the true spatial closure term. A more in-depth discussion of the choice of basis functions can be found in Section 2.4.

The basis coefficients of interest $\{\eta_{k}\}_{k=0}^M$ can again be estimated as part of the EnKF through the method of state augmentation. To do so, we will augment the state vector at time $t_i$ with the vector of coefficients $\boldsymbol \eta_{t_i}$ at time $t_i$ to form the new augmented state $\mathbf{z}_{t_i} = \begin{bmatrix}
    \mathbf{u}_{t_i} \\ \boldsymbol \eta_{t_i}
\end{bmatrix}$. If the true observation process is described by equation (\ref{eq:obsmodel}), then we can let $\mathbf{G}_{t_i} = \begin{bmatrix}
    \mathbf{H}_{t_i} & \mathbf{0}_{N_{t_i} \times (M+1)}
\end{bmatrix}$ be our augmented linear observation matrix and obtain the equivalent observation model
\begin{align}
    \mathbf{y}_{t_i} &= \mathbf{G}_{t_i} \mathbf{z}_{t_i} + \boldsymbol \epsilon_{t_i}, \quad
    \boldsymbol \epsilon_{t_i} \sim \mathcal{N}(\mathbf{0},\mathbf{R}_{t_i}).
\end{align}
The discrete-time evolution model $\mathcal{M}^z_{t_i}(\mathbf{z}_{t_i})$ for the augmented state may be found by solving the system of stochastic differential equations
\begin{equation}
\label{eq:pde-sde}
\begin{aligned}
        \partial_t u&= \mathcal{F}(u) + \sum_{k=0}^M \sigma_k \eta_k b_k, \\
        \text{d} \eta_0 &= \text{d}W_{0},\\
        \text{d} \eta_1 &= \text{d}W_{1},\\
        \vdots \\
        \text{d} \eta_M&= \text{d}W_{M}.\\
\end{aligned}
\end{equation}
Samples from the approximate solution to (\ref{eq:pde-sde}) can be obtained by e.g. applying central differences and using a stochastic solver such as the Euler-Maruyama method.

Given the augmented observation model and time-discrete evolution model, the ensemble Kalman filter provides approximations for the forecast and analysis distributions of the coefficients $\{\eta_k(t)\}_{k=0}^M$, which induce forecast and analysis distributions for the closure term through equation (\ref{eq:forecast_closure}).

\subsection{Choice of spatial basis functions for spatial closure term estimation}

\begin{figure}[H]
\centering
        \begin{subfigure}{0.45 \linewidth}
                \centering
                \includegraphics[scale = 0.25]{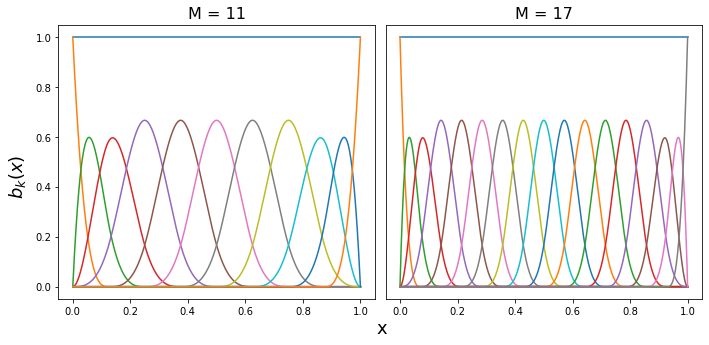}
            \caption{Two cubic B-spline bases over $\Omega = [0,1]$.}
            \label{fig:1DBspline}
        \end{subfigure} \quad 
        \begin{subfigure}{0.45 \linewidth}
            \centering
                    \includegraphics[scale = 0.25]{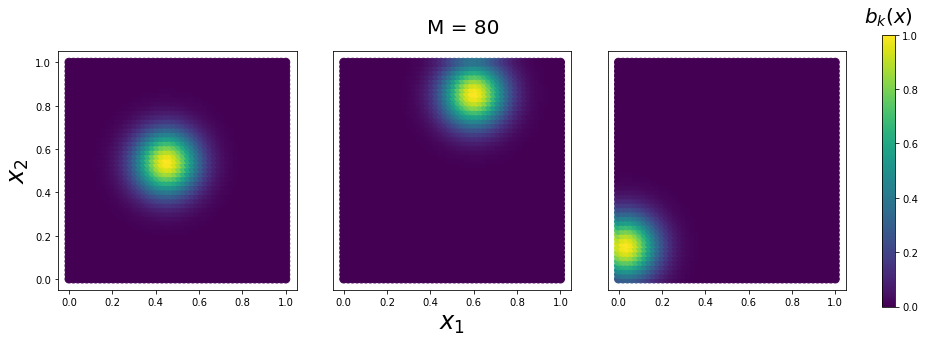}
            \caption{Three Gaussian basis functions over $\Omega = [0,1]^2$.}
            \label{fig:2DRBF}
    \end{subfigure}
    \caption{Examples of spatial basis functions.}
    \label{fig:1}
\end{figure}

The choice of basis functions should be determined by the practitioner's knowledge of the system and belief about the missing physics. For example, the user may wish to enforce periodicity with a Fourier basis or discontinuities with splines or step functions. For our numerical experiments, we have found that a suitable default choice of basis in one spatial dimension is the cubic B-Spline basis, which generates functions that are continuous in their first two spatial derivatives. This modeling choice produces estimates that are visually quite smooth. An example of a cubic B-spline basis is given in Figure \ref{fig:1DBspline}. The splines near the boundary have small support and therefore induce a large posterior variance when the observations are equally spaced and have identically distributed errors; however, they may be necessary to ensure the basis has sufficient flexibility to estimate the function in this region. This issue of increased bias and variance of nonparametric function estimators near the boundary of the domain is common in the nonparametric function estimation literature \cite{KYUNGJOON1998289}. The bases presented in Figure \ref{fig:1DBspline} have equally spaced knots, but if there is prior knowledge on the behavior of the closure term, the knots can be concentrated in areas where the closure term is known to vary rapidly in space.

In higher dimensions, we have found success with a radial basis function (RBF) basis using Gaussian kernels,
\begin{equation}
    b_k(\mathbf{x}) = e^{-\frac{||\mathbf{x} - \mathbf{c}_k||^2_2}{h^2}}, \quad k = 1, \ldots, M.
\end{equation}
This basis typically performs well on standard multivariate function approximation problems and has the universal approximation property, though it 
may produce estimates that are visually slightly "bumpy" owing to the peaks of the individual Gaussian kernels. We follow \cite{du2006radial} and select the centroids $\{\mathbf{c}_k
\}_{k=0}^M$ by k-means clustering over the discretized spatial mesh. We have similarly found success by following \cite{lowe1988multivariable} and setting the bandwidth parameter to
\begin{equation}
    h = \frac{d_{max}}{N_x},
\end{equation}
where $d_{max} := \underset{i,j}{max} \: ||\mathbf{c}_i - \mathbf{c}_j||$ is the maximum pairwise distance between the centroids. Figure \ref{fig:2DRBF} displays three Gaussian kernels from a set of $M = 80$ basis functions. The spatial mesh for clustering was chosen to be a two-dimensional regular grid over a rectangular domain. The Gaussian kernel basis lacks a natural way of modeling the closure term on the boundary, but the size of the kernels offers a nice balance between imposing smoothness on the closure term estimate and modeling local behavior.
A major limitation of the current methodology is the lack of regularization for the basis expansion coefficients. Thus, bases with high frequency components that are typically used in conjunction with hard-thresholding, such as wavelet or polynomial bases, may be unsuitable in the current framework. However, smooth bases with local or approximately local support, such as B-splines or Gaussian RBFs, appear to work quite well in our contexts.

\subsection{Estimation of EnKF hyperparameters with maximum likelihood}
Our method requires the specification of the number of basis functions $M$, as well as the scale parameters $\{\sigma_k\}_{k=0}^M$ that control the degree of variability in our forecast model for the closure term. Larger values of $\sigma_k$ and $M$ lead to more uncertainty in our analysis distribution for $\xi$ and potentially unstable estimates, while smaller values of $\sigma_k$ and $M$ may cause our estimates of $\xi$ to be rigid and unresponsive to the data. These hyperparameters can be difficult to specify a priori, as the time-evolution of the closure term depends on both the state and the unknown relationship between the state and closure. Furthermore, estimation of scale and correlation parameters through state augmentation can often fail owing to nonlinear relationships between the parameters and the data \cite{stroud2007sequential}. An alternative approach to estimation via state augmentation is the method of maximum marginal likelihood, also known as empirical Bayes or evidence maximization. However, for many nonlinear systems the marginal likelihood is intractable and must be approximated. A common approach in the EnKF literature is to approximate the marginal distribution of the data with a Gaussian distribution \cite{Stroud:2010,AnAdaptiveEnsembleKalmanFilter}, which can be derived mathematically when interpreting the EnKF as approximating the forecast distribution itself with a Gaussian distribution. Suppose we have auxiliary parameters $\boldsymbol \theta$, e.g. $\boldsymbol \theta = (M, \sigma_0,\sigma_1,\ldots,\sigma_M)$, that we wish to estimate using maximum marginal likelihood. If we let $\hat{\mathbf{m}}^f_{t_i,\boldsymbol \theta}$ be the estimated forecast mean and $\hat{\mathbf{P}}^f_{t_i,\boldsymbol \theta}$ be the estimated forecast covariance for a given value of $\boldsymbol \theta$, we can write our optimization problem as
\begin{align}
\label{eq:optproblem}
    \boldsymbol \theta_{MML} &= \underset{\boldsymbol \theta}{\text{argmax}} \:\: \text{log } p(\mathbf{y}_{t_1}, \mathbf{y}_{t_2}, \ldots, \mathbf{y}_{t_{N_o}} \mid \boldsymbol \theta) \\
    &= \underset{\boldsymbol \theta}{\text{argmax}} \sum_{i=1}^{N_o} \text{log } p(\mathbf{y}_{t_i} \mid \mathbf{y}_{t_1}, \mathbf{y}_{t_2}, \ldots, \mathbf{y}_{t_{i-1}}, \boldsymbol \theta) \\
    &\approx \underset{\boldsymbol \theta}{\text{argmin}} \sum_{i=1}^{N_o} [\text{det}(\mathbf{H}_{t_i}\hat{\mathbf{P}}^f_{t_i,\boldsymbol \theta}\mathbf{H}_{t_i}^T) + (\mathbf{y}_{t_i} - \hat{\mathbf{m}}^f_{t_i,\boldsymbol \theta})^T (\mathbf{H}_{t_i}\hat{\mathbf{P}}^f_{t_i,\boldsymbol \theta}\mathbf{H}_{t_i}^T)^{-1} (\mathbf{y}_{t_i} - \hat{\mathbf{m}}^f_{t_i,\boldsymbol \theta})]. \label{eq:optfunction}
\end{align}
The approximation in the last step is a result of both the Gaussian approximation for the marginal likelihood and Monte Carlo error in the estimated forecast means and covariances. 

Maximum marginal likelihood turns the estimation of $\boldsymbol \theta$ into a stochastic optimization problem. Each evaluation of the objective function requires a complete run of the ensemble Kalman filter to estimate the forecast mean $\hat{\mathbf{m}}^f_{t_i,\boldsymbol \theta}$ and covariance $\hat{\mathbf{P}}^f_{t_i,\boldsymbol \theta}$. These mean and covariance estimates are random due to finite ensemble sizes and the probabilistic nature of the measurement perturbations, Brownian motion sample paths, and random initializations of the ensemble members. Hence, the objective function is also random and must be treated with care.

Though running the entire filter on a large-scale system may be computationally expensive, the expense of the objective function could be mitigated by the use of specialized optimization routines such as Bayesian Optimization, which can be modified to accommodate noise in the objective function \cite{frazier2018tutorialbayesianoptimization}. In this work, we take a crude approach by using large ensemble sizes, fixing the stochastic elements of the optimization function, and performing grid search optimization as if the function were deterministic. Another potential pitfall of maximum marginal likelihood is that it does not necessarily guarantee the best estimate of the closure term, as the loss function is focused directly on the state-space predictions. However, our numerical experiments suggest promising results with maximum marginal likelihood as an estimation method for the auxiliary parameters. An investigation into the relationship between the marginal likelihood and the error of the EnKF estimates will be presented in Section 3. Future work could be dedicated to developing more computationally or statistically efficient methods for hyperparameter estimation.



\section{Numerical Experiments}
In this section, we illustrate the performance of our method using numerical experiments. In the examples presented, we learn the closure term from data simulated via the Fisher-KPP reaction-diffusion equation and advection-diffusion equation. For the reaction-diffusion equation, we examine the model under partial misspecification of the reaction term. The incomplete model for the advection-diffusion equation is examined under completely missing physics due to the omission of the advection term. We study both equations in both one and two spatial dimensions, as the simple, one-dimensional case is advantageous for visualization and the relative computational efficiency makes it an ideal setting for parametric studies, while the two-dimensional case showcases the method's performance under additional complexity. For the one-dimensional case, we perform a parametric study to determine the sensitivity of our method to the choice of a common scale parameter and the number of basis functions. We also investigate the influence of the prior distribution on our method, as well as the method's efficacy under two experimental settings: 1) low signal-to-noise ratio and spatially dense data, and 2) high signal-to-noise ratio and spatially sparse data. In the two-dimensional case, we simply report estimates after mild parameter tuning for brevity.

\subsection{Fisher-KPP equation}
Our first example is the problem of closure estimation for a system governed by the Fisher-KPP reaction-diffusion equation \cite{fisher1937wave,kolmogoroff1988study}, which has both a linear diffusion term and logistic reaction term
\begin{equation}
    \label{eq:complete-kpp}
    \partial_t u = D \nabla^2 u + u\Big(1-\frac{u}{K} \Big).
\end{equation}
This system has two model parameters: $K$, which controls the carrying capacity of the population density $u$, and the diffusivity $D$, which controls the speed of spatial spread for that population. The Fisher-KPP equation has enjoyed widespread use in the modeling of invasive populations and traveling waves and has diverse applications in cell biology, ecology, and combustion theory \cite{el2019revisiting}. We chose to use this equation in our example as it is a simple, well-studied nonlinear PDE that is commonly applied to real-world problems.

Suppose we lack full knowledge of the reaction term. Using the closure modeling framework, we can write our governing equation as
\begin{equation}
    \label{eq:incomplete-kpp}
    \partial_t u = D\nabla^2  u +u(1- u) + \phi.
\end{equation}
We will take the unknown closure term to be $\phi = (1-\frac{1}{K}) u^2$, which happens to correspond to a misspecified carrying capacity in equation \ref{eq:complete-kpp}. The nonlinear nature of this closure term makes for an interesting case-study, as the ensemble Kalman filter analysis updates are linear in the observations. We will estimate this closure term for the Fisher-KPP equation under two experimental settings for the i.i.d. Gaussian measurements, investigate the influence of the prior for the initial closure term on subsequent closure estimates, and perform a parametric study to determine the sensitivity of the closure term estimates to the choice of method hyperparameters.

\begin{figure}[H]
    \begin{center}
    \includegraphics[scale=0.45]{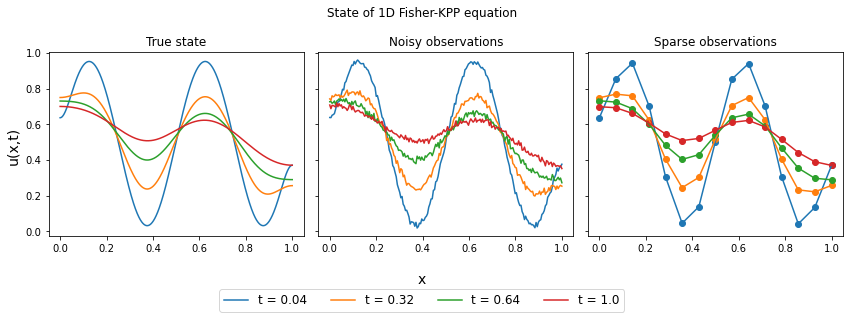}
    \end{center}
    \vspace{-0.2in}
    \caption{True value of the state (left) compared to spatially dense, high noise observations (middle) and spatially sparse, high signal-to-noise ratio observations (right). The state and observations are plotted at four distinct time points, which are denoted by different colors.}
    \label{fig:1d kpp sol}
\end{figure}

\begin{figure}[H]
    \begin{center}
        \begin{tabular}{lccc}
            \includegraphics[scale=0.4]{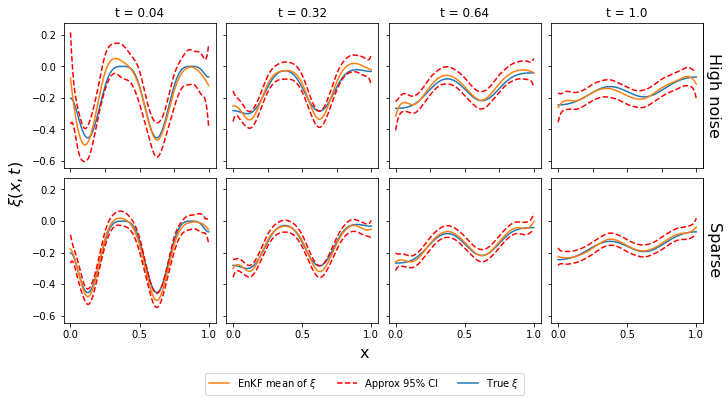}
        \end{tabular}
    \end{center}
    \vspace{-0.2in}
    \caption{Estimated closure term using the high noise measurements (top) and sparse measurements (bottom). The EnKF hyperparameters used are $(\sigma,M,N_{ens}) = (0.1,11,1000)$.}
    \label{fig:est1Dkpp}
\end{figure}

In all of the one-dimensional numerical experiments, we solve the system over an equally-spaced discretization of the spatial domain $\Omega = [0,1]$ with $N_x = 1000$ grid points. We use Neumann boundary conditions $\partial_x u(0,t) = \partial_x u(1,t) = 0$ and initial condition $u(x,0) = \text{sin}(4 \pi x) + 1$. Figure \ref{fig:1d kpp sol} shows the solution of the system when $(D,K) = (0.01,2/3)$ as it evolves over time. These model parameters were chosen so that the missing closure term would have a moderate but notable effect on the system dynamics. We can see that the solution is attracted towards the steady state value at $u = 2/3$, rather than the theorized $u = 1$ from the incomplete model (\ref{eq:incomplete-kpp}). Figure \ref{fig:1d kpp sol} also shows observations of the state under two conditions: spatially dense with high noise $(N_{t_i},\gamma) = (200,0.01)$ and spatially sparse with low noise $(N_{t_i},\gamma) = (15,0.001)$. For the sparse observation case, the plot of linearly interpolated observations is overlaid with a scatterplot of the observations to aid in visualization.
For estimating the unknown closure term, we assume equal Brownian motion scale parameters $\sigma_0 = \sigma_1 = \ldots = \sigma_M =: \sigma$ for the basis expansion coefficients to limit the dimension of the hyperparameter optimization problem. For the one-dimensional examples, we also use the spline basis shown in Figure \ref{fig:1DBspline} as our choice of spatial basis functions. Estimates of the missing closure term with parameters chosen by grid-search optimization can be observed in Figure \ref{fig:est1Dkpp}. The observations used for these estimates were taken at $N_o = 25$ time points $t_i = i/25$. For each time point shown in the figure, the ensemble Kalman filter analysis mean and approximate 95\% credible intervals directly after an analysis update are plotted against the true closure term. The credible intervals are centered at the EnKF mean with half-widths of $z_{0.975}$ times the EnKF standard deviation, where $z_{0.975}$ is the $97.5$-th percentile of the standard normal distribution. Notice that our method is able to track the evolution of the missing term over time with reasonable posterior coverage: for all but two of the scenarios shown, the true closure term is entirely contained by the pointwise credible intervals. The two remaining cases are the noisy setting at $t=0.32$ and dense setting at $t=0.04$, for which the pointwise credible intervals capture the true closure term at 99\% and 91\% of the spatial grid points respectively. While our EnKF analysis means do not exactly match the true closure terms, they are reasonable estimates given the data noise and sparsity, as well as the approximate nature of the data assimilation algorithm.

\begin{figure}[H]
    \begin{center}
        \begin{tabular}{lccc}
        \includegraphics[scale=0.3]{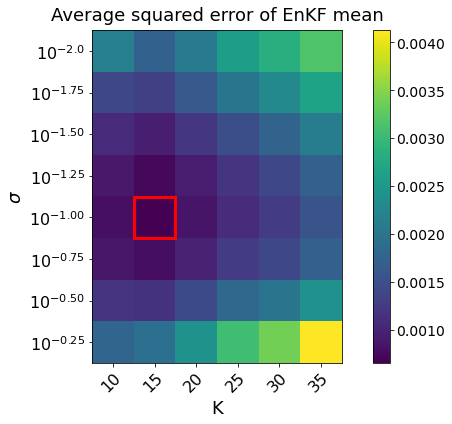}
        \includegraphics[scale=0.3]{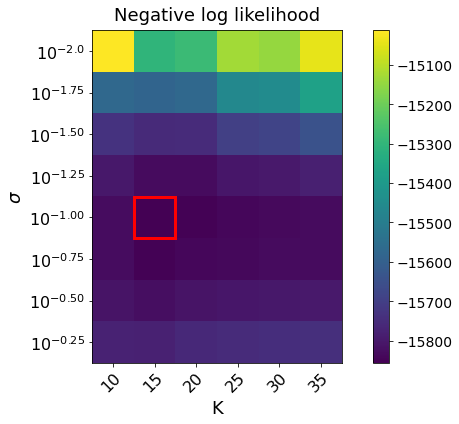}
        \includegraphics[scale=0.3]{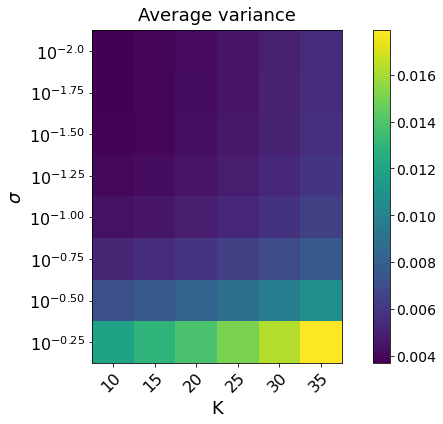}
        \end{tabular}
    \end{center}
    \vspace{-0.2in}
    \caption{Comparison of the normalized L2-error of the EnKF mean as an estimator of the true closure (left) and the negative marginal log-likelihood of the data (right) as functions of the scale parameter $\sigma$ and number of basis functions $M$.}
    \label{fig:1dKPP grid search}
\end{figure}

\begin{figure}[H]
 \begin{center}
         \includegraphics[scale=0.35]{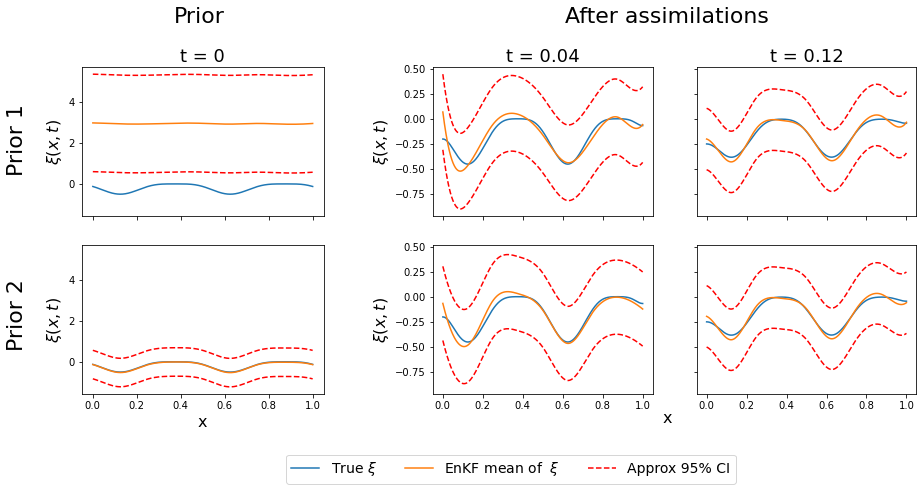}
 \end{center}
 \vspace{-0.2in}
 \caption{Comparison of $
         \xi$ estimation without (top) and with (bottom) high density near the true value in the prior of $\xi(\mathbf{x},0)$. We show the initial prior distribution (left), as well as the EnKF estimates after one (middle) and three (right) analysis updates are completed. Note the y-axis is changed after observations are taken.}
     \label{fig:badIC}
\end{figure}

To determine the sensitivity of our method to the choice of spatial basis and common scale parameter $\sigma$, as well as assess the relationship between the marginal likelihood and error of the closure term estimates, a parametric study is provided in Figure \ref{fig:1dKPP grid search}. The measurements were taken in the spatially dense, noisy setting with $N_o = 25$. In this figure, we plot the negative of the (approximate) log likelihood given in (\ref{eq:optfunction}), as well as a discrete approximation to the normalized $L^2$ error of the EnKF mean $\hat{\xi}$, which is defined to be
\begin{equation}
    \text{StandardizedL2Error} = \sqrt \frac{\iint(\hat{\xi}-\xi)^2\text{d}\mathbf{x}\text{dt}}{\iint(\xi-\iint\xi \: \text{d}\mathbf{x} \text{dt})^2\text{d}\mathbf{x}\text{dt}},
\end{equation}
which is a functions of $\sigma$ and the number of splines $M$. $L^2$ loss is a common choice for error assessment in machine learning models, which we elected to normalize due to the small scale of the closure term. We see that the minimizers of both the $L^2$ error and negative log-likelihood loss surfaces coincide over the coarse grid, suggesting that the negative log-likelihood is a reasonable loss function for closure term estimation.
Another matter of interest is the influence of the prior distribution for the spatial basis coefficients of the closure term at the initial time on the subsequent forecast and analysis estimates of the closure term. This is of particular import, as the functional form of the closure term may be completely unknown. In Figure \ref{fig:badIC}, the estimates obtained from utilizing a prior that places low density on the closure term at the initial time are contrasted with estimates obtained from a prior that is centered around coefficients which minimize the squared distance from the true closure term. After assimilating three observations under the spatially dense, high noise experimental setting, the differences are nearly visually imperceptible. As this result suggests the estimates are not sensitive to the choice of prior, for all other numerical experiments we use prior distributions that are centered around the least-squares optimal closure term at the initial time.

\begin{figure}[H]
    \begin{center}
        \includegraphics[scale=0.35]{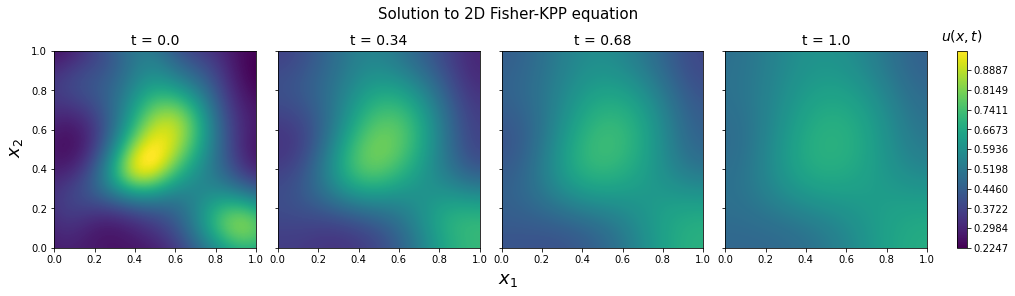}
    \end{center}
    \vspace{-0.2in}
    \caption{True solution of the 2D Fisher-KPP equation at four selected
    time points.}
    \label{fig:2d kpp sol}
\end{figure}

\begin{figure}[H]
    \begin{center}
        \includegraphics[scale=0.25]{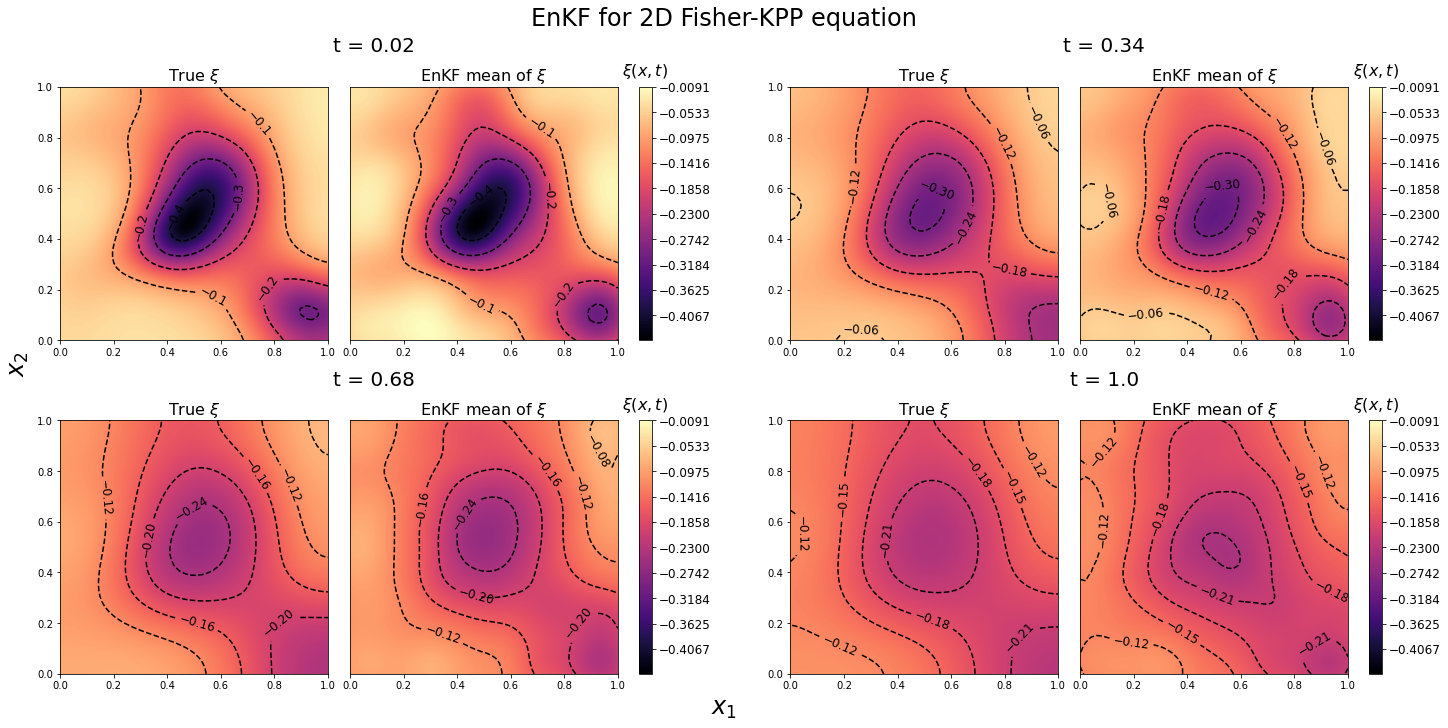}
    \end{center}
    \vspace{-0.2in}
    \caption{Contour plots for analysis mean estimates of the closure term for the 2D Fisher-KPP equation when $M = 80, N_{ens} = 1000$ and $\sigma = 0.025$.}
    \label{fig:2DKPPest}
\end{figure}

\begin{figure}[H]
    \begin{center}
        \begin{tabular}{lccc}
            \includegraphics[scale=0.35]{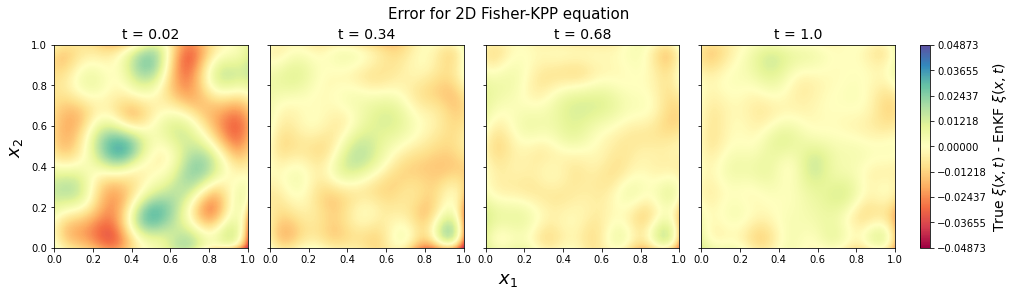}
        \end{tabular}
    \end{center}
    \vspace{-0.2in}
    \caption{Residual error of the analysis mean from the ground truth closure term at four different time points for the 2D Fisher-KPP equation when $M = 80, N_{ens} = 1000$, and $\sigma = 0.025$.}
    \label{fig:2DKPPerror}
\end{figure}

To demonstrate the performance of our method in higher spatial dimensions, we also present estimates of the closure term in the two-dimensional case. The true state at four selected time points with model parameters $(D,K) = (0.01,2/3)$ is given in Figure \ref{fig:2d kpp sol}. The system was solved via forward-time, central-space finite differences over a $50 \times 50$ discretization of the spatial domain $\Omega = [0,1]^2$. The boundary condition used is $\frac{\partial u}{\partial \mathbf{n}}(\mathbf{x},t) = 0$ for all $\mathbf{x} \in \partial \Omega$, where $\frac{\partial u}{\partial \mathbf{n}}$ is the normal derivative of $u$. The initial condition was generated from a scaled, exponentiated Gaussian process with squared-exponential covariance and is assumed to be known and fixed.
Figure \ref{fig:2DKPPest} shows contour plots for the analysis mean estimates of the closure term at four specific time points. These estimates were obtained using $N_o = 50$ temporal observations at $t_i = i/50$ with experimental settings $(N_{t_i},\gamma) = (2500,0.0025)$. We see that the EnKF mean is capable of tracking the evolution of the closure term, as there are only small errors visible from the contours and the shape of the closure term is largely preserved. Error plots are given in Figure \ref{fig:2DKPPerror} that show the difference between the true closure term and the EnKF analysis mean estimate at four time points. Besides a slight underestimation of the closure at the lower right-hand corner, there does not seem to be any obvious systematic deviations in our estimates, and the method seems to perform quite well.

\subsection{Advection-diffusion equation}
Our next physics-based model exemplar is the advection-diffusion equation
\begin{equation}
    \partial_t u = D \nabla^2 u -   \nabla \cdot (c  u).
\end{equation}
This system has two model parameters: the diffusivity $D$, which controls the rate of diffusion, and velocity $c$, which controls the rate of advection. The advection-diffusion equation has been used to model the transport of various quantities, including mass, energy, or heat, in diverse fields such as hydrology, chemical engineering, atmospheric science, and biology \cite{DEHGHAN20045}. For example, the advection-diffusion equation has been used to model the dispersion of pollutants in rivers and the forced cooling of windings in turbo generators by passing fluids \cite{DEHGHAN20045}.

Suppose our incomplete knowledge of the system is
\begin{equation}
    \partial_t u = D\nabla^2 u + \phi.
\end{equation}
Then $\phi = -\nabla \cdot  (c u)$ is our closure term, which we will again estimate using our proposed methodology. In this example, we have completely missing physics in our incomplete understanding of the system. We specify a Gaussian peak initial condition for this system in all experiments, which will enforce interesting spatial heterogeneity in the closure term. Specifically, some regions of space will have numerically zero closure term values, while other regions experience quite rapid temporal changes at early points in the trajectory. As such, one might expect the spatial heterogeneity of this closure term to pose a particular challenge when a shared scale parameter is used for all basis coefficients. As in the previous example, in the one-dimensional case we estimate the closure term using two experimental conditions: 1) high noise and spatially dense observations, and 2) low noise and spatially sparse observations. We furthermore perform a similar parametric study on the sensitivity of our estimates to the number of spatial basis functions $M$ and shared scale parameter $\sigma$. Specific to the advection-diffusion exemplar, we also apply our method in the case of severe model specification by considering an advection-dominated problem, for which we report results in both one and two spatial dimensions.

\begin{figure}[H]
    \begin{center}
    \includegraphics[scale=0.4]{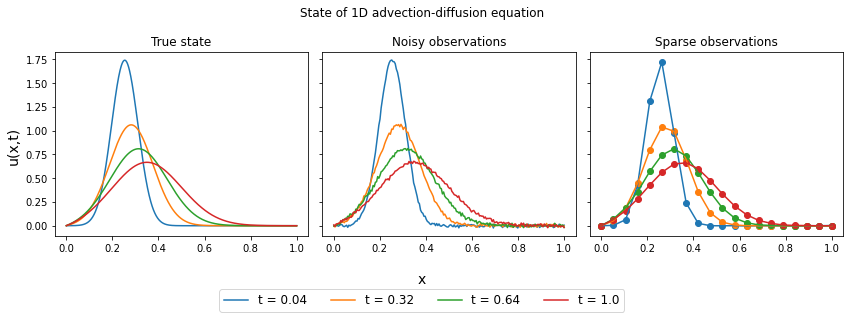}
    \end{center}
    \vspace{-0.2in}
    \caption{True value of the state (left) compared to dense, noisy observations (middle) and sparse, clean observations (right).}
    \label{fig:1d ade sol}
\end{figure}

In all one-dimensional numerical experiments, we solve this system over an equally-spaced discretization of the spatial grid $\Omega = [0,L]$ with $N_x = 1000$ spatial grid points. We use Dirichlet boundary condition $u(0,t) = u(L,t) = 0$ and initial condition $u(x,0) = e^{-200(x-0.25)^2}$. The solution of the complete system with parameters $(c,D) = (0.1,0.01)$ is shown in Figure \ref{fig:1d ade sol}, together with measurements under spatially dense with high noise $(N_{t_i},\gamma) = (200,0.01)$ and spatially sparse with low noise $(N_{t_i},\gamma ) = (20,0.001)$ experimental settings. Slightly more spatial observations were used in the sparse setting for the advection-diffusion equation than the Fisher-KPP equation, as there are large differences in the spatial observations near the Gaussian peak.

\begin{figure}[H]
    \begin{center}
        \begin{tabular}{lccc}
            \includegraphics[scale=0.45]{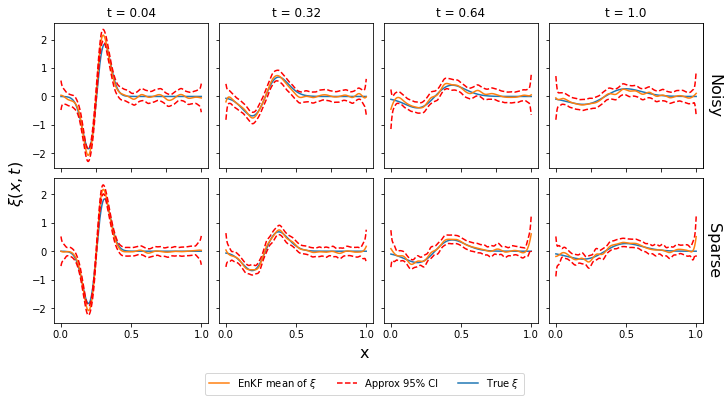}
        \end{tabular}
    \end{center}
    \vspace{-0.2in}
    \caption{Estimated closure term in the $c = 0.1$ case using the high noise measurements (top) and sparse measurements (bottom). The EnKF hyperparameters used are $(\sigma,M,N_{ens}) = (0.1,11,1000)$.}
        \label{fig:1d ade est}
\end{figure}

Analysis mean estimates of the closure term in the $c = 0.01$ case are given in figure \ref{fig:1d ade est}. Like the Fisher-KPP equation, a spline basis and shared scale parameter is used for the prior model. The closure estimates were obtained with $N_o = 25$ temporal observations at $t_i = i/25$ and the hyperparameter estimates were obtained by grid search. The pointewise 95\% credible intervals contain the true closure at 91.5\% of the spatial grid points at time $t=0.04$ in the dense setting. For the sparse setting, the 95\% credible intervals contain the closure at 87.5\% of the spatial grid points at $t=0.04$ and 99.1\% of the spatial grid points at $t = 0.64$. For both settings, the estimated closure term appears to have high error near the peaks of the function at $t=0.04$, which are in spatial regions that experience rapid changes at early time points in the trajectory. This suggests an underestimation of $\sigma_k$ for the bases in this region (due to the assumption of a shared $\sigma$ parameter) or possibly a need to have time-varying scale coefficients.

\begin{figure}[H]
    \begin{center}
        \begin{tabular}{lccc}
            \includegraphics[scale=0.3]{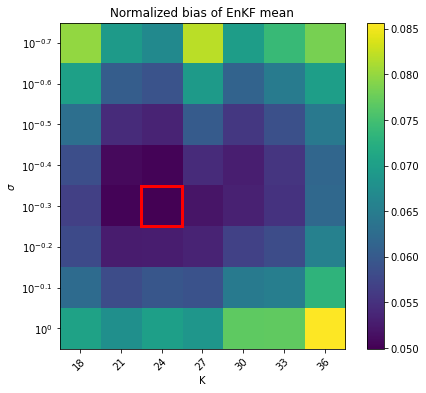}
            & \includegraphics[scale=0.3]{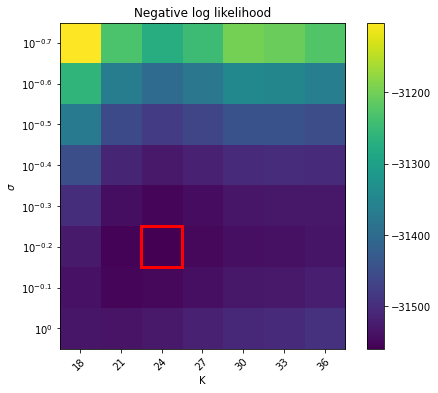}
        \end{tabular}
    \end{center}
    \vspace{-0.2in}
    \caption{Comparison of the bias of the EnKF mean as an estimator of the true closure (left) and the marginal likelihood of the data (right) as functions of the scale parameter $\sigma$ and number of basis functions $M$.}
    \label{fig:1dADE grid search}
\end{figure}

Another parametric study was performed in Figure \ref{fig:1dADE grid search} that supports the findings from Figure \ref{fig:1dKPP grid search}. Once again, the observations were collected in the spatially dense, high noise setting with $N_o = 25$. From the figure, we see the optimal $\sigma$ value in terms of log-likelihood is close to, though does not quite match, the optimal value in terms of normalized $L^2$ error, while the optimal number of bases $M$ is the same in both cases. Once again, the log-likelihood appears to be a reasonable criterion for hyperparameter selection.

\begin{figure}[H]
    \begin{center}
        \includegraphics[scale=0.35]{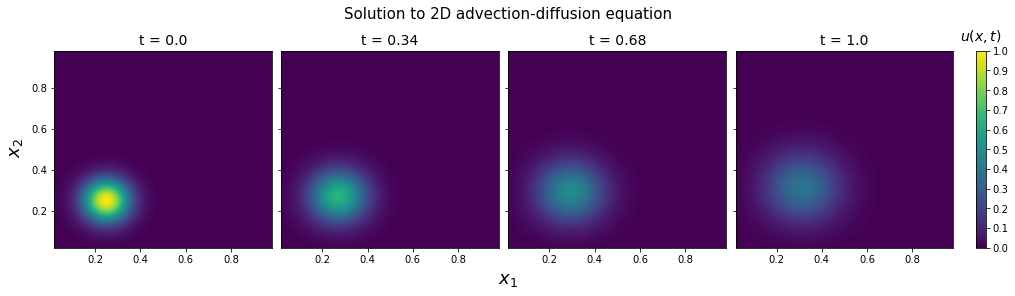}
    \end{center}
    \vspace{-0.2in}
    \caption{True solution of the 2D advection-diffusion equation at four selected time points.}
    \label{fig:2d advdiff}
\end{figure}

We also consider the advection-diffusion equation in two spatial dimensions. For model parameters $(c,D) = (0.1,0.01)$, we solve the system with solved via forward-time, central-space finite differences over a $50 \times 50$ discretization of the spatial domain $\Omega = [0,1]^2$. The boundary condition used is $u(\mathbf{x},t) = 0$ for all $\mathbf{x} \in \partial \Omega$, and the initial condition is a Gaussian peak $u(\mathbf{x},0) = e^{-100||\mathbf{x}-0.25\mathbf{1}_2||^2_2}$. The solution to the system at four time point is given in Figure \ref{fig:2d advdiff}.

\begin{figure}[H]
     \begin{center}
        \includegraphics[scale=0.3]{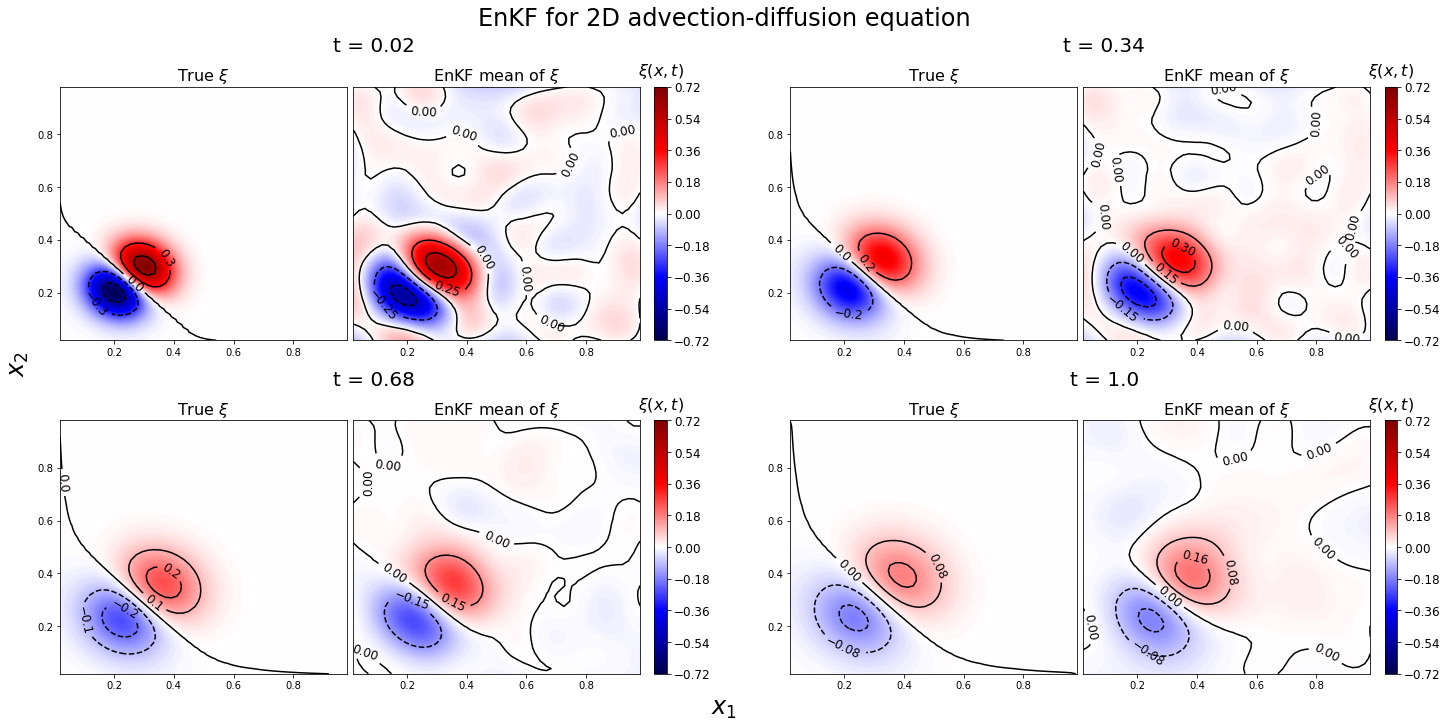}
     \end{center}
     \vspace{-0.2in}
     \caption{Contour plots for estimates of the closure term for the 2D advection-diffusion equation when $M = 81, N_{ens} = 1000$ and $\sigma = 0.1$.}
     \label{fig:2d advdiff est}
\end{figure}

\begin{figure}[H]
    \begin{center}
        \begin{tabular}{lccc}
            \includegraphics[scale=0.35]{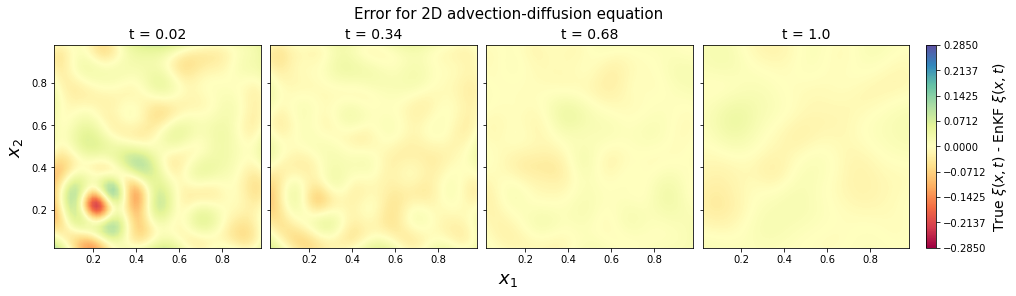}
        \end{tabular}
    \end{center}
    \vspace{-0.2in}
    \caption{Residual error for the closure term of the 2D advection-diffusion equation at four different time points when $\gamma = 0.0025, M = 80, N_{obs} = 50$ and $\sigma = 0.1$.}
    \label{fig:2d advdiff error}
\end{figure}

The EnKF analysis mean estimates of the closure term are compared with the true closure term in Figure \ref{fig:2d advdiff est}. The analysis estimates were computed using $N_o$ observations were taken at time points $t_i = i/50$ with $(N_{t_i},\gamma) = (2500,0.0025)$. The analysis mean estimates appear to be noisier than the estimates for the Fisher-KPP equation, with visible small-scale deviations present throughout space at each analysis step presented. However, the closure term estimates appear to track the true closure term quite well at the areas of high magnitude. Error plots from Figure \ref{fig:2d advdiff error} tell a slightly different story. These plots demonstrate that the error of the analysis mean is highest near the Gaussian peak after the first analysis step. This suggests an underestimation of the scale parameters $\sigma_k$ for the bases in that spatial region, which may be caused by the assumption of a shared $\sigma$ value for all basis coefficients.


\begin{figure}[H]
    \begin{center}
    \includegraphics[scale=0.5]{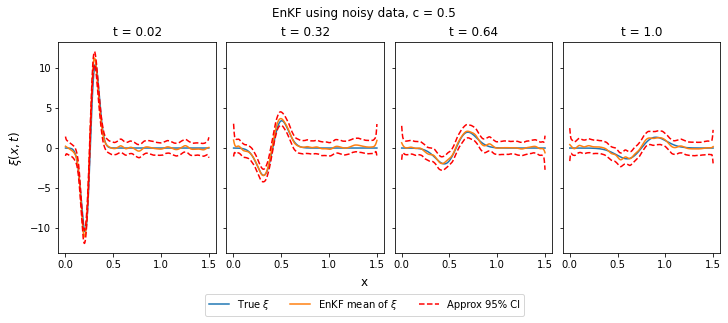} 
    \end{center}
    \begin{center}
            \includegraphics[scale=0.3]{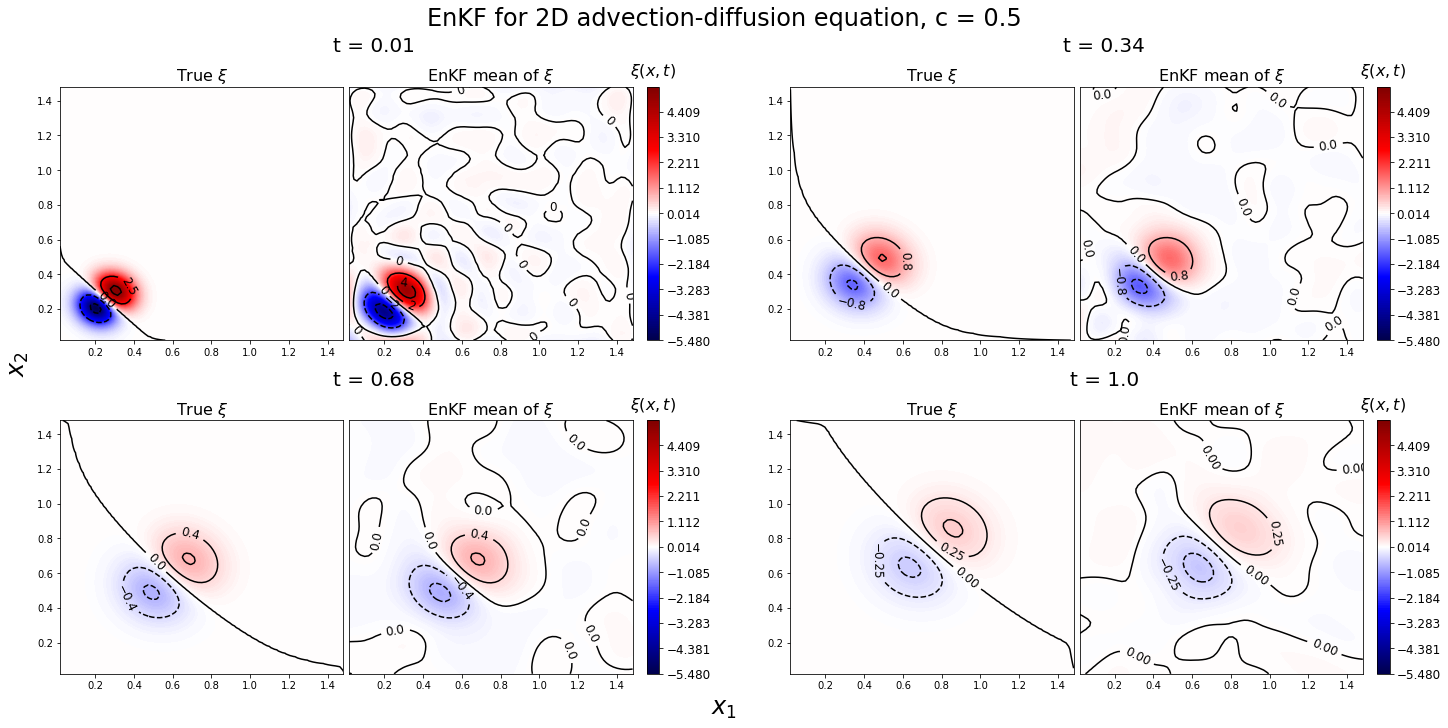}
    \end{center}
    \vspace{-0.2in}
    \caption{Estimates for the closure term of the advection-diffusion example when $c=0.5$.}
    \label{fig:2d-advdiff-advdominant}
\end{figure}

A potential concern in the previous examples is that the problem is not advection-dominant and therefore may not be viewed as significantly different from the reaction-diffusion example. To demonstrate that our method remains functional even in the advection-dominant case, we increase the advection coefficient from $c=0.1$ to $c=0.5$, and since the increase in advection would cause the solution to approach the boundary of the previous domain, consider the new domain $\Omega = [0,1.5]^d$ for $d=1,2$. We also double the temporal sampling rate and increase the spatial grid size and spatial sampling rate by a multiplicative factor of $1.5$. The EnKF estimates are shown in Figure \ref{fig:2d-advdiff-advdominant}. The estimates appear reasonable, indicating that the success of the method is not limited to a narrow range of system dynamics.

\section{Conclusion}

In this study, we have introduced a novel methodology for estimating closure terms in spatiotemporal models of dynamical systems, leveraging the ensemble Kalman filter (EnKF) framework. Our approach addresses the significant challenges associated with closure modeling, particularly in the context of partial differential equations (PDEs) where traditional methods often fall short due to computational demands or assumptions of noise-free observations.

Through extensive numerical experiments involving the Fisher-KPP reaction-diffusion equation and the advection-diffusion equation, we demonstrated the efficacy of our method in accurately estimating closure terms even under conditions of sparse and noisy data. The incorporation of basis expansion techniques not only enhanced computational efficiency but also provided a structured means to capture the spatial dynamics of the closure terms, leading to improved uncertainty quantification.

Our findings indicate that the proposed two-stage approach, which decouples closure term estimation from subsequent machine learning modeling, offers significant advantages in terms of interpretability and computational resource management. The results suggest that our method is robust across different experimental settings, including varying signal-to-noise ratios and spatial configurations.

Future work will focus on refining the hyperparameter estimation process and exploring the integration of advanced machine learning techniques to further enhance the predictive capabilities of our closure modeling framework. Additionally, we aim to extend our methodology to more complex dynamical systems and investigate its applicability in real-world scenarios, thereby contributing to the broader field of scientific machine learning and its applications in the natural sciences and engineering.

\section*{Acknowledgments}
Sandia National Laboratories is a multimission laboratory managed and operated by National Technology and Engineering Solutions of Sandia , LLC., a wholly owned subsidiary of Honeywell International, Inc., for the U.S. Department of Energy’s National Nuclear Security Administration under contract DE-NA-0003525. This paper describes objective technical results and analysis. Any subjective views or opinions that might be expressed in the paper do not necessarily represent the views of the U.S. Department of Energy or the United States Government. 

This article has been authored by an employee of National Technology \& Engineering Solutions of Sandia, LLC under Contract No. DE-NA0003525 with the U.S. Department of Energy (DOE). The employee owns all right, title and interest in and to the article and is solely responsible for its contents. The United States Government retains and the publisher, by accepting the article for publication, acknowledges that the United States Government retains a non-exclusive, paid-up, irrevocable, world-wide license to publish or reproduce the published form of this article or allow others to do so, for United States Government purposes. The DOE will provide public access to these results of federally sponsored research in accordance with the DOE Public Access Plan https://www.energy.gov/downloads/doe-public-access-plan.

\nocite{*}
\bibliographystyle{unsrt}
\bibliography{Draft}

@book{evensen2006data,
  title={Data Assimilation: The Ensemble Kalman Filter},
  author={Evensen, G.},
  isbn={9783540383017},
  url={https://books.google.com/books?id=VJ2oOecHhOYC},
  year={2006},
  publisher={Springer Berlin Heidelberg}
}

@misc{rackauckas2021universaldifferentialequationsscientific,
      title={Universal Differential Equations for Scientific Machine Learning}, 
      author={Christopher Rackauckas and Yingbo Ma and Julius Martensen and Collin Warner and Kirill Zubov and Rohit Supekar and Dominic Skinner and Ali Ramadhan and Alan Edelman},
      year={2021},
      eprint={2001.04385},
      archivePrefix={arXiv},
      primaryClass={cs.LG},
      url={https://arxiv.org/abs/2001.04385}, 
}

@article{PDEfind,
author = {Samuel H. Rudy  and Steven L. Brunton  and Joshua L. Proctor  and J. Nathan Kutz },
title = {Data-driven discovery of partial differential equations},
journal = {Science Advances},
volume = {3},
number = {4},
pages = {e1602614},
year = {2017},
doi = {10.1126/sciadv.1602614},
URL = {https://www.science.org/doi/abs/10.1126/sciadv.1602614},
eprint = {https://www.science.org/doi/pdf/10.1126/sciadv.1602614}
}

@book{GPR,
    author = {Rasmussen, Carl Edward and Williams, Christopher K. I.},
    title = {Gaussian Processes for Machine Learning},
    publisher = {The MIT Press},
    year = {2005},
    month = {11},
    doi = {10.7551/mitpress/3206.001.0001},
    url = {https://doi.org/10.7551/mitpress/3206.001.0001},
    eprint = {https://direct.mit.edu/book-pdf/2514321/book\_9780262256834.pdf},
}

@article{evensen2009ensemble,
  title={The ensemble Kalman filter for combined state and parameter estimation},
  author={Evensen, Geir},
  journal={IEEE Control Systems Magazine},
  volume={29},
  number={3},
  pages={83--104},
  year={2009},
  publisher={IEEE}
}

@article{WIKLE20071,
title = {A Bayesian tutorial for data assimilation},
journal = {Physica D: Nonlinear Phenomena},
volume = {230},
number = {1},
pages = {1-16},
year = {2007},
note = {Data Assimilation},
issn = {0167-2789},
doi = {https://doi.org/10.1016/j.physd.2006.09.017},
url = {https://www.sciencedirect.com/science/article/pii/S016727890600354X},
author = {Christopher K. Wikle and L. Mark Berliner}
}

@article{Yang_2021,
   title={B-PINNs: Bayesian physics-informed neural networks for forward and inverse PDE problems with noisy data},
   volume={425},
   ISSN={0021-9991},
   url={http://dx.doi.org/10.1016/j.jcp.2020.109913},
   DOI={10.1016/j.jcp.2020.109913},
   journal={Journal of Computational Physics},
   publisher={Elsevier BV},
   author={Yang, Liu and Meng, Xuhui and Karniadakis, George Em},
   year={2021},
   month=jan, pages={109913} }

@article{Brunton_2016,
   title={Discovering governing equations from data by sparse identification of nonlinear dynamical systems},
   volume={113},
   ISSN={1091-6490},
   url={http://dx.doi.org/10.1073/pnas.1517384113},
   DOI={10.1073/pnas.1517384113},
   number={15},
   journal={Proceedings of the National Academy of Sciences},
   publisher={Proceedings of the National Academy of Sciences},
   author={Brunton, Steven L. and Proctor, Joshua L. and Kutz, J. Nathan},
   year={2016},
   month=mar, pages={3932–3937} 
}

@article{Kennedy,
author = {Kennedy, Marc C. and O'Hagan, Anthony},
title = {Bayesian calibration of computer models},
journal = {Journal of the Royal Statistical Society: Series B (Statistical Methodology)},
volume = {63},
number = {3},
pages = {425-464},
doi = {https://doi.org/10.1111/1467-9868.00294},
url = {https://rss.onlinelibrary.wiley.com/doi/abs/10.1111/1467-9868.00294},
eprint = {https://rss.onlinelibrary.wiley.com/doi/pdf/10.1111/1467-9868.00294},
year = {2001}
}

@article{RAISSI2019686,
title = {Physics-informed neural networks: A deep learning framework for solving forward and inverse problems involving nonlinear partial differential equations},
journal = {Journal of Computational Physics},
volume = {378},
pages = {686-707},
year = {2019},
issn = {0021-9991},
doi = {https://doi.org/10.1016/j.jcp.2018.10.045},
url = {https://www.sciencedirect.com/science/article/pii/S0021999118307125},
author = {M. Raissi and P. Perdikaris and G.E. Karniadakis}
}

@conference{osti_1875381,
  author       = {Acquesta, Erin},
  title        = {Adapting Verification and Validation Principles to a Credibility Process for Scientific Machine Learning.},
  url          = {https://www.osti.gov/biblgio/1875381},
  place        = {United States},
  organization = {Sandia National Lab. (SNL-NM), Albuquerque, NM (United States)},
  year         = {2021},
  month        = {06}
  }

@misc{ebers2023discrepancymodelingframeworklearning,
      title={Discrepancy Modeling Framework: Learning missing physics, modeling systematic residuals, and disambiguating between deterministic and random effects}, 
      author={Megan R. Ebers and Katherine M. Steele and J. Nathan Kutz},
      year={2023},
      eprint={2203.05164},
      archivePrefix={arXiv},
      primaryClass={stat.ML},
      url={https://arxiv.org/abs/2203.05164}, 
}

@article{DMD,
author = {Wu, Ziyou  and Brunton, Steven L.  and Revzen, Shai },
title = {Challenges in dynamic mode decomposition},
journal = {Journal of The Royal Society Interface},
volume = {18},
number = {185},
pages = {20210686},
year = {2021},
doi = {10.1098/rsif.2021.0686},

URL = {https://royalsocietypublishing.org/doi/abs/10.1098/rsif.2021.0686},
eprint = {https://royalsocietypublishing.org/doi/pdf/10.1098/rsif.2021.0686}
}

@article{du2006radial,
  title={Radial basis function networks},
  author={Du, K -L and Swamy, MNS},
  journal={Neural networks in a softcomputing framework},
  pages={251--294},
  year={2006},
  publisher={Springer}
}

@article{park1991universal,
  title={Universal approximation using radial-basis-function networks},
  author={Park, Jooyoung and Sandberg, Irwin W},
  journal={Neural computation},
  volume={3},
  number={2},
  pages={246--257},
  year={1991},
  publisher={MIT Press}
}

@article{lowe1988multivariable,
  title={Multivariable functional interpolation and adaptive networks},
  author={Lowe, David and Broomhead, D},
  journal={Complex systems},
  volume={2},
  number={3},
  pages={321--355},
  year={1988}
}

@book{goldstein2007bayes,
  title={Bayes linear statistics: Theory and methods},
  author={Goldstein, Michael and Wooff, David},
  year={2007},
  publisher={John Wiley \& Sons}
}

@article{hartigan1969linear,
  title={Linear bayesian methods},
  author={Hartigan, JA},
  journal={Journal of the Royal Statistical Society: Series B (Methodological)},
  volume={31},
  number={3},
  pages={446--454},
  year={1969},
  publisher={Wiley Online Library}
}

@book{west2006bayesian,
  title={Bayesian forecasting and dynamic models},
  author={West, Mike and Harrison, Jeff},
  year={2006},
  publisher={Springer Science \& Business Media}
}

@article{katzfuss2016understanding,
  title={Understanding the ensemble Kalman filter},
  author={Katzfuss, Matthias and Stroud, Jonathan R and Wikle, Christopher K},
  journal={The American Statistician},
  volume={70},
  number={4},
  pages={350--357},
  year={2016},
  publisher={Taylor \& Francis}
}

@article{Stroud:2010,
   author    =  "J. R. Stroud and M. L. Stein and B. M. Lesht and D. J. Schwab and D. Beletsky",
   title     =  "An ensemble Kalman filter and smoother for satellite data assimilation",
   year      =  "2010",
   journal   =  "J. Amer. Stat. Assoc.",
   volume    =  "105",
   pages     =  "978--990"
}

@article{Anderson:2001,
   author    =  "J. L. Anderson",
   title     =  "An ensemble adjustment Kalman filter for data assimilation",
   year      =  "2001",
   journal   =  "Mon. Wea. Rev.",
   volume    =  "129",
   pages     =  "2884--2903"
}

@book{hastie2009elements,
  title={The elements of statistical learning: data mining, inference, and prediction},
  author={Hastie, Trevor and Tibshirani, Robert and Friedman, Jerome H and Friedman, Jerome H},
  volume={2},
  year={2009},
  publisher={Springer}
}

@article{buhmann2000radial,
  title={Radial basis functions},
  author={Buhmann, Martin Dietrich},
  journal={Acta numerica},
  volume={9},
  pages={1--38},
  year={2000},
  publisher={Cambridge university press}
}

@book{de1978practical,
  title={A practical guide to splines},
  author={De Boor, Carl and De Boor, Carl},
  volume={27},
  year={1978},
  publisher={springer New York}
}

@book{ramsay2002applied,
  title={Applied functional data analysis: methods and case studies},
  author={Ramsay, James O and Silverman, Bernard W},
  year={2002},
  publisher={Springer}
}

@article{subramanian2019error,
  title={Error estimation in coupled multi-physics models},
  author={Subramanian, Abhinav and Mahadevan, Sankaran},
  journal={Journal of Computational Physics},
  volume={395},
  pages={19--37},
  year={2019},
  publisher={Elsevier}
}

@inproceedings{neal2024investigating,
  title={Investigating Model Form Error Estimation for Sparse Data},
  author={Neal, Kyle D and Khalil, Mohammad and Portone, Teresa},
  booktitle={Annual Conference of the PHM Society},
  volume={16},
  number={1},
  year={2024}
}

@article{subramanian2023probabilistic,
  title={Probabilistic physics-informed machine learning for dynamic systems},
  author={Subramanian, Abhinav and Mahadevan, Sankaran},
  journal={Reliability Engineering \& System Safety},
  volume={230},
  pages={108899},
  year={2023},
  publisher={Elsevier}
}

@misc{stevenshaas2024learningnonlineardynamicsusing,
      title={Learning Nonlinear Dynamics Using Kalman Smoothing}, 
      author={Jacob Stevens-Haas and Yash Bhangale and Aleksandr Aravkin and Nathan Kutz},
      year={2024},
      eprint={2405.03154},
      archivePrefix={arXiv},
      primaryClass={math.DS},
      url={https://arxiv.org/abs/2405.03154}, 
}

@book{sarkka2019applied,
  title={Applied stochastic differential equations},
  author={S{\"a}rkk{\"a}, Simo and Solin, Arno},
  volume={10},
  year={2019},
  publisher={Cambridge University Press}
}

@incollection{kolmogoroff1988study,
  title={Study of the diffusion equation with growth of the quantity of matter and its application to a biology problem},
  author={Kolmogoroff, A and Petrovsky, I and Piscounoff, N},
  booktitle={Dynamics of curved fronts},
  pages={105--130},
  year={1988},
  publisher={Elsevier}
}

@article{fisher1937wave,
  title={The wave of advance of advantageous genes},
  author={Fisher, Ronald Aylmer},
  journal={Annals of eugenics},
  volume={7},
  number={4},
  pages={355--369},
  year={1937},
  publisher={Wiley Online Library}
}

@article{kalman1960new,
  title={A new approach to linear filtering and prediction problems},
  author={Kalman, Rudolph Emil},
  year={1960}
}

@misc{north2022bayesianapproachspatiotemporaldatadriven,
      title={A Bayesian Approach for Spatio-Temporal Data-Driven Dynamic Equation Discovery}, 
      author={Joshua S. North and Christopher K. Wikle and Erin M. Schliep},
      year={2022},
      eprint={2209.02750},
      archivePrefix={arXiv},
      primaryClass={stat.ME},
      url={https://arxiv.org/abs/2209.02750}, 
}

@article{KASHYAP2024111474,
title = {A Gaussian-process assisted model-form error estimation in multiple-degrees-of-freedom systems},
journal = {Mechanical Systems and Signal Processing},
volume = {216},
pages = {111474},
year = {2024},
issn = {0888-3270},
doi = {https://doi.org/10.1016/j.ymssp.2024.111474},
url = {https://www.sciencedirect.com/science/article/pii/S0888327024003728},
author = {Sahil Kashyap and Timothy J. Rogers and Rajdip Nayek}
}

@article{garg2022physics,
  title={Physics-integrated hybrid framework for model form error identification in nonlinear dynamical systems},
  author={Garg, Shailesh and Chakraborty, Souvik and Hazra, Budhaditya},
  journal={Mechanical Systems and Signal Processing},
  volume={173},
  pages={109039},
  year={2022},
  publisher={Elsevier}
}

@article{schmid2010dynamic,
  title={Dynamic mode decomposition of numerical and experimental data},
  author={Schmid, Peter J},
  journal={Journal of fluid mechanics},
  volume={656},
  pages={5--28},
  year={2010},
  publisher={Cambridge University Press}
}

@article{stroud2007sequential,
  title={Sequential state and variance estimation within the ensemble Kalman filter},
  author={Stroud, Jonathan R and Bengtsson, Thomas},
  journal={Monthly weather review},
  volume={135},
  number={9},
  pages={3194--3208},
  year={2007}
}

@article{fruhwirth1994data,
  title={Data augmentation and dynamic linear models},
  author={Fr{\"u}hwirth-Schnatter, Sylvia},
  journal={Journal of time series analysis},
  volume={15},
  number={2},
  pages={183--202},
  year={1994},
  publisher={Wiley Online Library}
}

@article{hornik1991approximation,
  title={Approximation capabilities of multilayer feedforward networks},
  author={Hornik, Kurt},
  journal={Neural networks},
  volume={4},
  number={2},
  pages={251--257},
  year={1991},
  publisher={Elsevier}
}

@article{cybenko1989approximation,
  title={Approximation by superpositions of a sigmoidal function},
  author={Cybenko, George},
  journal={Mathematics of control, signals and systems},
  volume={2},
  number={4},
  pages={303--314},
  year={1989},
  publisher={Springer}
}

@article{burgers1998analysis,
  title={Analysis scheme in the ensemble Kalman filter},
  author={Burgers, Gerrit and Jan van Leeuwen, Peter and Evensen, Geir},
  journal={Monthly weather review},
  volume={126},
  number={6},
  pages={1719--1724},
  year={1998}
}

@article{godsill2004monte,
  title={Monte Carlo smoothing for nonlinear time series},
  author={Godsill, Simon J and Doucet, Arnaud and West, Mike},
  journal={Journal of the american statistical association},
  volume={99},
  number={465},
  pages={156--168},
  year={2004},
  publisher={Taylor \& Francis}
}

@article{el2019revisiting,
  title={Revisiting the Fisher--Kolmogorov--Petrovsky--Piskunov equation to interpret the spreading--extinction dichotomy},
  author={El-Hachem, Maud and McCue, Scott W and Jin, Wang and Du, Yihong and Simpson, Matthew J},
  journal={Proceedings of the Royal Society A},
  volume={475},
  number={2229},
  pages={20190378},
  year={2019},
  publisher={The Royal Society Publishing}
}

@article{kingma2014adam,
  title={Adam: A method for stochastic optimization},
  author={Kingma, Diederik P and Ba, Jimmy},
  journal={arXiv preprint arXiv:1412.6980},
  year={2014}
}

@article{liu1989limited,
  title={On the limited memory BFGS method for large scale optimization},
  author={Liu, Dong C and Nocedal, Jorge},
  journal={Mathematical programming},
  volume={45},
  number={1},
  pages={503--528},
  year={1989},
  publisher={Springer}
}

@article{houtekamer1998data,
  title={Data assimilation using an ensemble Kalman filter technique},
  author={Houtekamer, Peter L and Mitchell, Herschel L},
  journal={Monthly weather review},
  volume={126},
  number={3},
  pages={796--811},
  year={1998}
}

@article{lei2010comparison,
  title={Comparison of ensemble Kalman filters under non-Gaussianity},
  author={Lei, Jing and Bickel, Peter and Snyder, Chris},
  journal={Monthly Weather Review},
  volume={138},
  number={4},
  pages={1293--1306},
  year={2010}
}

@inproceedings{gillijns2006ensemble,
  title={What is the ensemble Kalman filter and how well does it work?},
  author={Gillijns, Steven and Mendoza, O Barrero and Chandrasekar, Jaganath and De Moor, BLR and Bernstein, Dennis S and Ridley, A},
  booktitle={2006 American control conference},
  pages={6--pp},
  year={2006},
  organization={IEEE}
}

@article {AssessingthePerformanceoftheEnsembleKalmanFilterforLandSurfaceDataAssimilation,
      author = "Yuhua Zhou and Dennis McLaughlin and Dara Entekhabi",
      title = "Assessing the Performance of the Ensemble Kalman Filter for Land Surface Data Assimilation",
      journal = "Monthly Weather Review",
      year = "2006",
      publisher = "American Meteorological Society",
      address = "Boston MA, USA",
      volume = "134",
      number = "8",
      doi = "10.1175/MWR3153.1",
      pages=      "2128 - 2142",
      url = "https://journals.ametsoc.org/view/journals/mwre/134/8/mwr3153.1.xml"
}

@inbook{Bengtsson_2008,
   title={Curse-of-dimensionality revisited: Collapse of the particle filter in very large scale systems},
   ISBN={0940600749},
   url={http://dx.doi.org/10.1214/193940307000000518},
   DOI={10.1214/193940307000000518},
   booktitle={Probability and Statistics: Essays in Honor of David A. Freedman},
   publisher={Institute of Mathematical Statistics},
   author={Bengtsson, Thomas and Bickel, Peter and Li, Bo},
   year={2008},
   pages={316–334} }

@article {AnAdaptiveEnsembleKalmanFilter,
      author = "Herschel L. Mitchell and P. L. Houtekamer",
      title = "An Adaptive Ensemble Kalman Filter",
      journal = "Monthly Weather Review",
      year = "2000",
      publisher = "American Meteorological Society",
      address = "Boston MA, USA",
      volume = "128",
      number = "2",
      doi = "10.1175/1520-0493(2000)128<0416:AAEKF>2.0.CO;2",
      pages=      "416 - 433",
      url = "https://journals.ametsoc.org/view/journals/mwre/128/2/1520-0493_2000_128_0416_aaekf_2.0.co_2.xml"
}

@misc{frazier2018tutorialbayesianoptimization,
      title={A Tutorial on Bayesian Optimization}, 
      author={Peter I. Frazier},
      year={2018},
      eprint={1807.02811},
      archivePrefix={arXiv},
      primaryClass={stat.ML},
      url={https://arxiv.org/abs/1807.02811}, 
}

@misc{sanderse2024scientificmachinelearningclosure,
      title={Scientific machine learning for closure models in multiscale problems: a review}, 
      author={Benjamin Sanderse and Panos Stinis and Romit Maulik and Shady E. Ahmed},
      year={2024},
      eprint={2403.02913},
      archivePrefix={arXiv},
      primaryClass={math.NA},
      url={https://arxiv.org/abs/2403.02913}, 
}

@article{DEHGHAN20045,
title = {Numerical solution of the three-dimensional advection–diffusion equation},
journal = {Applied Mathematics and Computation},
volume = {150},
number = {1},
pages = {5-19},
year = {2004},
issn = {0096-3003},
doi = {https://doi.org/10.1016/S0096-3003(03)00193-0},
url = {https://www.sciencedirect.com/science/article/pii/S0096300303001930},
author = {Mehdi Dehghan}
}

@book{sullivan2015introduction,
  title={Introduction to uncertainty quantification},
  author={Sullivan, Timothy John},
  volume={63},
  year={2015},
  publisher={Springer}
}

@article{KYUNGJOON1998289,
title = {Nonparametric kernel regression estimation near endpoints},
journal = {Journal of Statistical Planning and Inference},
volume = {66},
number = {2},
pages = {289-304},
year = {1998},
issn = {0378-3758},
doi = {https://doi.org/10.1016/S0378-3758(97)00082-7},
url = {https://www.sciencedirect.com/science/article/pii/S0378375897000827},
author = {Cha Kyung-Joon and William R. Schucany}
}

\end{document}